\documentstyle [12pt] {article}
\newcommand{\cs}[3]{{{#3} \brace {#1 #2}}}
\begin{document}
\begin{center}
{\huge Quantum Interference of Thermal Neutrons and Spin-Torsion Interaction }\\
\end{center}
\begin{center}
{\bf M.I.Wanas, M.Melek, and M.E.Kahil} 
\end{center}
\begin{center}
{\it Astronomy and Meteorology Department, Faculty of Science, Cairo University, 
Orman, Giza, Egypt.} 
\end{center}
\begin{center}
e.mail: Wanas@FRCU.EUN.EG
\end{center}

\begin{abstract}

The discrepancy of $0.8 \%$ between theory and the COW-experiment is interpreted.
This is done by using a new path equation other than the geodesic one. It is shown that this discrepancy
is possibly due to a type of interaction between the torsion, of space-time generated by
the background field, and the spin of the moving neutron. The results obtained are
discussed and compared with the experimental interpretation suggested by Arif et al. in 1994. 
As a byproduct, an upper limit is imposed on the free parameter of the new path equation used.   
     
\end{abstract}
\vskip 1truecm
\baselineskip= .9cm
\section{Introduction}     
 
~~~~~Colella, Overhauser and Werner suggested and carried out experiments
concerning the quantum interference of thermal neutrons [1], [2], [3]. This type
of experiments is known, in the literature, as the COW-experiment. The aim
of the experiment is to test the effect of the Earth's gravitational field on the
phase difference between two beams of thermal neutrons, one is more closer to
the Earth's surface than the other. The second version of the COW experiment [4] 
was carried out by Werner et al.(1988). This version is characterized by
a high accuracy in the measurements of the phase shift (1 part in 1000). The
measurements show that the experimental results are lower than the theoretical
calculations (using the Newtonian gravity) by about $0.8 \%$. This discrepancy
between theory and experiment has no satisfactory interpretation, so far.

It is suggested that the discrepancy may be due to non-Newtonian effects [4].
The possiblility, that this discrepancy may be due to some relativistic effects,
 can be easily ruled out. This is because relativistic effects are of the second order in the gravitational potential and will be much
smaller than $0.8 \%$.
 The question which emerges now is: 
 What is the origin of
this discrepancy? 
There are three possibilities to find out the origin of this discrepancy:\\
1. One of these possibilities is  to consider the origin of the discrepancy 
to be due to an experimental artifact [5],[6].\\ 
2. The second possibility concerns the basis upon which the theoretical prediction 
is made. \\
3. The discrepancy may have both theoretical and experimental origins. \\
However, a trial to explore the first possibility has been done recently [5], [6].
It is the aim of the present work to explore the second  and third possibilities, which can not 
be ruled out, in view of the recent results obtained by using the first possibility. 
The final word will be left for the third generation of the COW-experiment [6]. \\
\section{New Path Equation}
     In a trial to find possible paths, in Absolute Parallelism (AP) geometry [7],  
that can be considered as generalizations of geodesics of Riemannian geometry, 
the Bazanski's [8] approach is generalized.
We found that there are only three 
different paths of this type [9],
\begin{equation}
{{\frac{dJ^\mu}{dS^-}} + \cs{\nu}{\sigma}{\mu}\ J^\nu J^\sigma = 0}
\end{equation}
\begin{equation}
{{\frac{dW^\mu}{dS^o}} + \cs{\nu}{\sigma}{\mu}\ W^\nu W^\sigma = 
- {\frac{1}{2}}\Lambda_{(\nu \sigma)} . ^\mu ~ W^\nu~ W^\sigma}
\end{equation}    
 \begin{equation}
{{\frac{dV^\mu}{dS^+}} + \cs{\nu}{\sigma}{\mu}\ V^\nu V^\sigma = 
- \Lambda_{(\nu \sigma)} . ^\mu ~ V^\nu~ V^\sigma}
\end{equation}      
where $\cs{\nu}{\sigma}{\mu}$ is the Christoffel
symbol, $\Lambda^\mu_{. \nu \sigma}$ is the torsion of the AP-space, $S^-$, $S^o$ and $S^+$ are the parameters varying along these paths respectively. $J$, $W$ and $V$ are the tangents to these paths respectively. 
      
 The common feature in the set of equations of paths
is the appearance of a term containing the torsion of space-time. The vanishing
of this term reduces the equations to that of a geodesic (or null-geodesic
after reparameterization). The attracting feature in the 3-new equations is 
that the coefficients  of the torsion term  are $0, {\frac{1}{2}}, 1$ respectively, i. e.
there is a jump of step ${\frac{1}{2}}$ among the coefficients. This urges to generalize
the affine connections of the AP-geometry to get the family of paths in which the
coefficient of the torsion term jumps by steps of ${\frac{1}{2}}$. The equation describing
this family is found to be in the form [10]:

\begin{equation}
{{\frac{dZ^\mu}{d\tau}} + \cs{\nu}{\sigma}{\mu}\ Z^\nu Z^\sigma = 
- {\frac{n}{2}} \alpha \gamma \Lambda_{(\nu \sigma)} . ^\mu~ Z^\nu~ Z^\sigma}
\end{equation}    
where       
\begin{equation}        
{Z^\mu Z_\mu = Z^2 ,}
\end{equation}  
$Z^\mu$ is the tangent to the path, n is the natural number, $\alpha$ is the fine structure
constant and $\gamma$ is a numerical free parameter to be fixed. This equation has been 
interpreted as describing the motion of a spinning particle in a gravitational 
field. Consequently, the term on R.H.S. of (4) represents a type of interaction between 
the intrinsic spin of the moving test particle and the torsion of the background
gravitational field. We take $n = 0, 1, 2, ...$ for particles with spin $0, {\frac{1}{2}}, 1, ...$
respectively.

Assuming weak static background field and slowly moving test particle we found,
using equation (4), that the Newtonian potential $\Phi_N$ should be modified,
for spinning particles, to $\Phi_S$ given by the relation [10]:
\begin{equation}
{\Phi_S = (1 - {\frac{n}{2}} \alpha \gamma) \Phi_N .}  
\end{equation}

It is clear from (6) that for spinless particles (or macroscopic objects):
\begin{equation}
{\Phi_S = \Phi_N  ,}
\end{equation}
as we substitute $n=0$ in (6) .
\section{Theoretical Interpretation of the Discrepancy}
Now  one can use equation (4) to give an interpretation for the discrepancy in
the COW-experiment. In fact we are going to use the consequence of equation (4) given by equation (6)
since the following assumptions hold:\\
-thermal neutrons can be considered as $\underline{slowly}$ moving test particles, and \\
-the Earth's gravitational field can be considered as $\underline{weak}$ and 
$\underline{static}$. 

The phase difference $(\Delta \Omega)$ between the beams of neutrons in the 
COW-experiment is given by [11].
\begin{equation}
{(\Delta \Omega)_N = - {\frac{1}{\hbar}} \int^{ABD}_{ACD} \Phi_N dt ,}
\end{equation}
where ABD and ACD are the trajectories of the upper and lower beams of neutrons  in the
interferometer respectively . The index N is used to indicate that (8) is 
obtained using the Newtonian potential $\Phi_N$, and $\hbar$ is the Planck's 
constant. Since neutrons are spinning particles they will be affected by 
the torsion of space-time as suggested. Thus we replace $\Phi_N$ in (8) by $\Phi_S$ 
given by (6). In this case (8) will take the form :
\begin{equation}
{(\Delta \Omega)_S = - (1 - {\frac{n}{2}} \gamma \alpha) {\frac{1}{\hbar}} 
\int^{ABD}_{ACD} \Phi_N dt ,}
\end{equation}
i.e.,
\begin{equation}
{(\Delta \Omega)_S = (\Delta \Omega)_N - {\frac{n}{2}} \gamma \alpha 
(\Delta \Omega)_N .}
\end{equation}
The index S is used to indicate that (9) is obtained using the potential $\Phi_S$.
Taking the value of $\alpha = {\frac{1}{137}}$, $n = 1$ for spin ${\frac{1}{2}}$-
particles (neutrons), we easily get the following results:\\ 
(1) the theoretical value of the COW-experiment will decrease by about
$0.4 \%$ if we take $\gamma = 1$, \\
(2) the theoretical value will coincide with the experimental one if we take $\gamma = 2$. 
	   
\section{Discussion}
        The difference between theoretical predictions and experimental results of the COW-experiment [4] represents a real discrepancy, since it is eight times the sensitivity of the interferometer used. Now there are two possible different interpretations for this discrepancy. The first was suggested by Arif et al. in 1994 [6], in which it is suggested that this dicrepancy
may be accounted for by elastic deformation of the interferometer. However, they used the finite element model to calculate the hypothetical deformation of the crystal monolith. They stated that the accuracy of the finite element model is not sufficient to allow for systematic correction of such experiments. Also, the result given by those authors shows that the suggested systematic  effect forces the experiment to give values that exceed the original theoretical calculations by $0.37\%$; which is about 4-times the accuracy of the second version of this experiment. So we still have a discrepancy but with opposite sign. In place of having higher theoretical predications than experimental result by 8 parts in one thousand [4], we have now a lower 
 theoretical predications than experimental result by 4 parts in one thousand [6].

On the other hand the second possible interpretation of this discrepancy is given above in the present work. It suggests that this discrepancy may be accounted for by taking into consideration a suggested interaction between the spin of the neutrons and the torsion of the 
gravitational field. If we take $\gamma=2$ in equation (9), then theoretical predications coincide exactly (within the experimental error) with the experimental result. However, such suggested interaction needs much experimental efforts to be verified.

 From the above discussion, we have two possibilities for interpreting the discrepancy in the COW-experiment. It appears that one of them cannot rule out  the other. If we assume that this discrepancy is due to the first possibility alone, then one has to account for the $0.4\%$ increase of experimental result over the theoretical one (using the calculations given in [4]). On the other hand  if we assume that it is due to the second possibility alone, then if we take the parameter $\gamma=1$ one has to account for $0.4\%$ decrease of the experimental result below 
the theoretical one (using the calculation given in the present work); while if we take $\gamma=2$, the theoretical calculation will coincide with the experimental one. However, a  third possibility cannot be neglected. That is, elastic deformation of the interferometer
and spin-torsion interaction, both are responsibile for this discrepancy. In this case if we take in the present calculation $\gamma=1$ together with Arif et al. results [6], we get an exact coincidence between theoretical predications and experimental results of the COW-experiment.

As a byproduct an upper limit can be imposed on the parameter $\gamma$. This limit of this parameter is $\gamma=2$ if we take the second possibility alone to interpret the discrepancy. But if we
consider the third possibilty, the upper limit will be 1. However, the parameter $\gamma$ may
be less or much less than 1. Table(1) summarises  the results using the three possibilities. The final word of this discrepancy will come from the results of the third version of the COW-experiment. This might fix the sign and the upper limit of the numerical parameter $\gamma$ appearing on R.H.S. of equation (4).

\begin{center}
  Table 1: Comparison between the values of the discrepancy using the three possibilities     \end{center}            
\begin{center}        
\vspace{.5cm}
\begin{tabular}{|c|c|c|} \hline  
 ~~~~~~~~~Theory      & Original Calculation [4]& Spin-Torsion interaction \\ 
                      &                    & $\gamma=1$  ~~ $\gamma=2$  \\ 
  Experiment ~~~~~    &                    &                            \\
& & \\ \hline
& & \\
  COW[4]            &- 8                             &  -4   ~~~~~~~   0 \\ 
 & &  \\ \hline
& & \\
 COW+Finite && \\
 element model[6]   & + 4                             &  0   ~~~~~~  -4 \\ 
 & & \\ \hline 
\end{tabular}
\end{center}
{\it{The values of the discrepancy (= experiment -theory) are given as parts in one thousand.}}
       
\section*{References}
~~~~~[1]  A. W. Overhauser and R. Colella, Phys. Rev. Lett., {\bf 33}, 1237 (1974).

[2]  R. Colella, A. W. Overhauser, and S. A. Werner, Phys. Rev. Lett., {\bf 34}, 1472 (1975)

[3]  J. L. Staudenmann, S. A. Werner, R. Colella, and A. W. Overhauser, Phys. Rev.

~~~~~A, {\bf 21}, 1419 (1980).

[4]  S. A. Werner, H. Kaiser, M. Arif and R. Clothier, Physica B, {\bf 151}, 22 (1988).

[5]  H.P. Layer and G.L. Greene, Phys.lett. A, {\bf 155}, 450 (1991).

[6]  M. Arif, M.S. Dewey, G.L. Greene, D. Jacobson and S.A. Werner, Phys. Lett. A, 

~~~~~{\bf 185}, 154 (1994).

[7]  Recently attention is directed to the construction of field theories using 
AP-spaces 

~~~~~cf. F. I. Mikhail, and M. I. Wanas ,Proc. Roy. Soc. Lond., A{\bf 356}, 471 (1977);

~~~~~C. Moller, Mat. Fys. Medd. Dan. Vid. Selsk. {\bf 39}, 1 (1978);

~~~~~K. Hayashi and T. Shirafuji, Phys. Rev. D{\bf 60}, 176 (1979).

[8]  S. L. Bazanski, Ann. Inst. H. Poincar\'e, A{\bf 27}, 145 (1977).

[9]  M. I. Wanas, M. Melek and M. E. Kahil, Astrophys. Space Sci. 288, 273
(1995).

[10] Wanas, M. I. (1998) To appear in Astrophys.Space Sci.

[11]  D. M. Greenberger, Rev. Mod. Phys., {\bf 55}, 875 (1983).

\end{document}